\documentclass[epj]{svjour}

\usepackage{latexsym}
\usepackage{graphicx,float}

\begin{document}

\title{Fingering Instability in a Water-Sand Mixture}

\author{A. Lange\inst{1}, M. Schr\"oter\inst{2}, M. A. Scherer\inst{2},
A. Engel\inst{1} \and I. Rehberg\inst{2}
}

\institute{Institut f\"ur Theoretische Physik,
Otto-von-Guericke-Universit\"at, Postfach 4120,
D-39016 Magdeburg,
Germany
\and
Institut f\"ur Experimentelle Physik,
Otto-von-Guericke-Universit\"at,
Postfach 4120,
D-39016 Magdeburg,
Germany
}

\offprints{A. Lange}

\mail{A. Lange}

\date{Received: date / Revised version: date}

\abstract{The temporal evolution of a water-sand interface
driven by gravity is experimentally investigated. By means of a Fourier
analysis of the evolving interface the growth rates are determined for the
different modes appearing in the developing front. To model the observed
behavior we apply the idea of the Rayleigh-Taylor instability for two
stratified fluids. Carrying out a linear stability analysis we calculate
the growth rates from the corresponding dispersion relations for finite
and infinite cell sizes. Based on the theoretical results the viscosity
of the suspension is estimated to be approximately 100 times higher
than that of pure water, in agreement with other experimental findings.
} 

\maketitle

\section{Introduction}
\label{sec:1}
There has been great interest in the behavior of granular materials over the
last years (for a review see \cite{jaeger96} and references therein).
Examples for the surprising behavior of granular matter are
two-dimensional localized states, called oscillons
\cite{umbanhowar}, in vertically vibrated containers, stratification
phenomena observed while pouring granular mixtures onto a pile \cite{makse} or
singing sand audible in deserts \cite{goldsack}.

In contrast to the amount of phenomena one lacks a sound theoretical
explanation for the observations. Difficulties arise due to the highly
complex, disorded structure formed by the grains and their nonlinear internal
friction. Therefore, most theoretical approaches have been done
by numerical methods. Molecular-dynamics simulations and cellular automata
calculations are frequently applied, for recent reviews see
\cite{ristow}. These methods describe in some detail the
interactions between the particular grains \cite{buchholtz}.
Conclusions on the macroscopic behavior of granular matter are then drawn
from the simulations with a great number of particles. Despite the
recent progress in computer performance the size of the systems is still
too small for a quantitative comparison with real experiments
\cite{buchholtz,kohring}.

Hydrodynamic approaches to granular material are few
\cite{haff,jenkins,hayakawa,grossman}
and are associated with restrictions as no interparticle correlations
\cite{haff}, a Gaussian distribution in the 
velocity of the grains \cite{jenkins} or 
steady-state properties \cite{grossman}. Nevertheless, there are striking
phenomenological similarities in the observed patterns for pure granular
materials and pure fluids. Experiments with an inclined chute
\cite{pouliquen,huppert} or a vertically vibrated container \cite{melo94,fauve}
show the most noteable analogy. It is clear that granular media are different
from fluids but under certain conditions these differences are
{\it not prevailing}.

The flow of grains embedded in an interstitial fluid is dominated either by
the effects of grain inertia or by effects of fluid viscosity.
The Bagnold number $B$ \cite{bagnold} expresses the ratio of collision
forces between the grains to viscous forces in the fluid-grain mixture.
A small Bagnold number, $B < 40$, characterizes the regime of the
macro-viscous flow. In this regime the viscous interaction
with the pure fluid is important. Examples for this type of flow are mud
slides and the transport of water-sand mixtures in river beds.
At large Bagnold numbers, $B > 450$, the flow is called grain-inertia
regime where the grain-grain interactions dominate. All flows of grains
with air as interstitial fluid fall into the grain-grain regime.

Here, experiments are performed with sand dispersed in
water. The occuring shear rates, the mean particle diameters, and
the viscosity of water result in a Bagnold number of about $1$
\cite{schroeter}. This motivates the idea
to consider the water-sand mixture as fluid-like. In the experiments
we observe that the initial flat water-sand interface evolves into a
finger-like pattern. The measured velocity for the largest finger
was typically three times the Stokes velocity of a settling hard sphere.
Furthermore, the velocity was nearly independent of the mass of the sand and
the diameter of the particles \cite{schroeter}. This independence of particle
properties encourages a fluid-like description of the water-sand mixture.
To model this behavior we test a continuum approach which is based on a
well-known hydrodynamic instability.

The aim is to determine the critical parameters for the stability of the
pattern and the dispersion relation for the {\it whole} spectrum of
wave numbers detectable in the experiment. Particularly, the latter extends
significantly the objectives of earlier experiments with water and polysterene
spheres \cite{skjeltorp} and with water and glass beads \cite{didwania}.
The dependence of the initial wavelength of the developing pattern on the
width of the cell was studied in \cite{skjeltorp}. In \cite{didwania} the
experiments were focused on the evolution of voidage
shock fronts caused by a step increase or decrease in the fluidization
velocity for a fluidized bed. We, however, turn our attention to 
the temporal evolution of {\it all} wave numbers.

To model the water-sand mixture as a Newtonian fluid with effective
properties depending on the concentration of the particles is obviously a
simplifying description. In such a model it is assumed that the particle
concentration in the flowing mixture is almost constant. Furthermore, the
particles have to be large enough to neglect their Brownian diffusion.
Stability analyses with these simplifications were made for moderate
concentrated mixtures in horizontal Hagen-Poiseuille flow \cite{zhang}
and in sedimentation problems in inclined narrow channels
\cite{rubinstein,leung,amberg}. For the latter configuration even the
sediment was described by such a simple model and a fair agreement with
experimental results was found \cite{leung}. In the clear knowledge of
the limitations of a fluid model for the water-sand mixture we will
examine whether such an approach can catch the essence of the experimental
results.

In the following section, the experimental arrangement is described and the
results for the growth rates of the wave numbers are presented. In
Sect.~\ref{sec:3}
the model is explained and thereafter the calculated growth rates are compared
with those of the experiment (Sect.~\ref{sec:4}). The final section contains
our conclusions and some remarks about further prospectives.

\section{Experiment}
\label{sec:2}
\subsection {Experimental setup}
\label{sec:2.1}
A closed Hele-Shaw-like cell is used to investigate the
temporal evolution of a water-sand interface driven by gravity (see
Fig.\ \ref{setup}). The cell, a CCD-camera and a neon tube are fixed to a frame
which can be turned around a horizontal axis. This allows image analysis in the
comoving frame and ensures a homogeneous
illumination at every stage of the pattern forming process. The length of the
cell is 160 mm, the height
80 mm and the width 4 mm, respectively. The cell is filled with sand and
destilled water. As sand we use spherical glass particles (W\"{u}rth
Ballotini MGL) of different sizes and size distributions (see
Table\ \ref{table1} for details). Its material density is given by 2.45
g/cm$^{3}$. The rotation axes of the frame
is right beneath the sand layer. This minimizes the centrifugal forces on the
sand layer while the cell is turned. The cell is rotated by hand. To obtain
reproducible results the vertical and horizontal acceleration is measured by
acceleration sensors (ADXL05).
When the cell passes a rotation angle of
170 degree, a lightgate triggers a number of snapshots. This moment
defines the starting time of our measurements where we take images
every 20 ms for later analysis. The images have a dimension of 256 x 300
pixel. To achieve a reasonable resolution we only focus on a horizontal
length of 61 mm at the middle of the cell. This gives a resolution of 4.9
pixel/mm. After each rotation the suspension sediments until all particles
are settled. This process takes typically less than 1 minute. By means of
tracer particles it was proven that at this time the fluid in the cell is
at rest. The time between consecutive runs is at least 3 minutes to secure
independent runs. The influence of the waiting time on the occuring patterns
was tested and no effect concerning the wavelength of the fingers was observed.

\subsection {Experimental results}
\label{sec:2.2}
Fig.\ \ref{pattern} shows typical images of the sand-water cell at certain
stages. 20 ms after the series of snapshots is started, the initial flat sand
layer is modulated at small scales (Fig.\ \ref{pattern}(a)). These
disturbances are enhanced and give rise to sand fingers as seen in
Fig.\ \ref{pattern}(c). At later stages the fingers evolve to a mushroom-like
pattern (Fig.\ \ref{pattern}(d) and Fig.\ \ref{pattern}(e)). This type
of pattern was found also by numerical simulations for two stratified
suspensions of different concentration \cite{druzhinin}.

To analyze this behavior we apply a threshold algorithm to obtain the
water-sand interface. We look at every column of our digitized image to
determine the point where the grey scale exceeds a certain value. We start at
the bottom (water) and continue to the top (sand). In this way we track down the
interface of the pattern. Fig.\ \ref{temporal} shows the temporal evolution of
the images presented in Fig.\ \ref{pattern}. Here the interfaces of all
patterns are shown. While our detecting method works for the patterns during
the first stages, namely small scale modulations and sand fingers, it breaks
down for mushroom-like patterns. However, this is not crucial because those
patterns are beyond the scope of the linear analysis presented here.
Discrete-Fourier-Transformation (DFT) gives the Fourier
spectrum of each interface. Fig.\ \ref{fit} shows the temporal evolution of the
amplitude $A$ of a typical Fourier mode. It is seen that $A$ grows
exponentially from the first image to $t$ = 200 ms.
By an exponential fit
\begin{equation}
\label{efit}
A(k)=A_i(k)\exp(n(k)t)
\end{equation}
we obtain the growth rate $n$ for every wave number $k$ in our spectra, where
$A_i$ is the initial amplitude. For each fit it was monitored that $A(k)$ was
smaller than $40$ \% of its wavelength, the criterion for the linear regime
\cite{sharp}.

In order to test the reliability of our experimental setup we perform 100
independent runs with one set of material parameters. The particular values
of this set are given in the first row of Table I (Experiment I).
We only analyze image series where the angular velocity of
our rotating apparatus is larger than 6.6 rad/s. We find that 42 fast runs
show only a slight deviation in the angular velocity: 7.4 $\pm$ 0.1 rad/s.
These 42 measurements are analyzed to obtain a mean growth rate and a mean
initial amplitude for each wave number. The results are shown in
Fig.\ \ref{wavenumber}. It is seen that the growth rates starting with small
values increase with increasing wave number until they saturate at
larger $k$. In contrast the initial amplitude decreases for increasing
wave numbers. In the case of large wave numbers we do not obtain
{\it exponential}
fits for every experimental run. This is due to the fact that the amplitude
is very small and that we approach the limit of the resolution of our
image processing. The error bars in Fig.\ \ref{wavenumber} indicate that
the number of runs which can be analyzed decreases for larger wave numbers.

An obvious question concerning the underlying mechanism of the
pattern formation in our system is: How do different material
parameters effect the dispersion relation?
Therefore we carry out experiments with different material configurations
which are characterized in Table I. Using 2 g, 4 g, and 8 g of sand (Experiment
I, II, III) we observe a shift of all growth rates towards larger values
with increasing mass of sand. This effect is independent of the wave
number (Fig.\ \ref{material} (a)). Using three different size distributions
(Experiment I, IV, V) the ratio of the cell thickness of $4$ mm to the mean
particle diameter varies between $\sim 70$ and $\sim 40$.  As figure
\ref{material} (b) shows the mean particle diameter does not have any
significant influence on the dispersion relation $n(k)$. As a common feature
we find the same overall behavior for all material sets: The growth rates
increase for small $k$ and reach a plateau for larger values of $k$.

\section{Theory}
\label{sec:3}
We choose a two fluid system as a model to describe the
experimental results. In the initial state two
incompressible fluids of constant densities $\rho_1$ and
$\rho_2$ and constant dynamical viscosities $\mu_1$ and $\mu_2$ are arranged
in two horizontal strata. The index 1 (2) refers to the fluid at the bottom
(top) of the system. The pressure is a function of the
vertical coordinate $z$ only; $x$ and $y$ are the coordinates in the
plane perpendicular to $z$. The acceleration due to gravity acts in negative
$z$ direction. A {\it linear} stability analysis is carried out
for small disturbances of this initial state.
The instability of a planar interface $z=z_s(x, y)\equiv 0$ between
the two fluids is known as the Rayleigh-Taylor instability
\cite{sharp,rayleigh,chandra}.

We assume that the boundaries in $z$ direction are far from the
interface. Small changes $\delta z_s$ in the form of the interface cause a
pressure difference which is balanced by the product of the surface tension
$T_s$ and the curvature of the interface. Considering small disturbances
$\delta\rho$ in the density and $\delta p$ in the pressure the
Navier-Stokes equations read \cite{chandra}
\begin{eqnarray}
\label{1}
   \rho\partial_t u &=& -\partial_x\delta p
   +\mu\Delta u +\left( \partial_x w+\partial_z u\right)\partial_z\mu\quad,\\
\label{2}
  \rho\partial_t v &=& -\partial_y\delta p
  +\mu\Delta v +\left( \partial_y w+\partial_z v\right)\partial_z\mu\quad,\\
\label{3}
  \rho\partial_t w &=& -\partial_z\delta p
  +\mu\Delta w +2\partial_z w\partial_z\mu-g\delta\rho\nonumber\\
  &&+T_s\left(\partial_x^2+\partial_y^2\right)
  \delta z_s\;\delta (z-z_s)\quad ,
\end{eqnarray}
where $\partial_i =\partial/\partial i$, $i=x$, $y$, $z$, $t$.
The $z$ dependence of $\mu$ gives rise to the third term at the right-hand
side of (\ref{1}-\ref{3}) since the viscous part in the Navier-Stokes equations
is $\partial_j\left[ \mu\left( \partial_j v_i +\partial_i v_j
\right)\right]$ for an incompressible fluid. For convenience, we adhere to
$\partial_z \mu$ though $\partial_z \mu$ is different from zero only at the
interface. The components of the velocity field $\vec{v}$ are $v_x=u$, $v_y=v$,
and $v_z=w$ and are considered small, so that Eqs.\ (\ref{1}-\ref{3}) contain
only terms which are linear in the disturbances. The delta-function
$\delta (z-z_s)$ ensures that the surface tension appears at the
interface $z_s$ between the two fluids. The equation of continuity (mass
conservation) for an incompressible fluid is
\begin{equation}\label{4}
{\rm div}\;\vec{v}=\partial_x u+\partial_y v+\partial_z w=0\quad .
\end{equation}
Additionally the equation
\begin{equation}
\label{5}
  \partial_t\delta \rho = -(\vec{v}{\rm grad} )\rho = -w\partial_z \rho
\end{equation}
relates the temporal variations in the density fluctuations to the density
jump at the interface which moves with $w$ in $z$ direction.
The Eqs.\ (\ref{1}-\ref{5}) govern the linearized system. The disturbances are
analyzed into {\it normal modes} thus seeking solutions which $x$ and $t$
dependence is proportional to $\exp (ik x+nt)$. The wave number is denoted by
$k$ and $n(k)$ is the growth rate of the corresponding mode $k$. If the fluid
is confined between two rigid planes the boundary conditions are
\begin{equation}
\label{6}
  w=\partial_z w=0\qquad {\rm  at~~}z=\pm\infty\quad .
\end{equation}
where we shift the planes to infinity for the sake of simplicity. The other
boundary conditions are related to the interface. All three components of the
velocity and the tangential viscous stresses must be continuous. Using the
exponential ansatz for the continuous velocity components, Eq.\ (\ref{4})
gives the continuity of $\partial_z w$, too. The continuity of
$\mu (\partial^2_z +k^2)w$ is the condition which ensures that the two
tangential stress components are continuous across the interface. Inserting
the exponential ansatz into (\ref{1}-\ref{5}) and determining the solution of
$w$ in each region in such a way that the boundary condition (\ref{6}) as
well as the interfacial conditions are satisfied 
leads to the dispersion relation \cite{chandra}
\begin{eqnarray}
\label{11}
&&-\biggr\{{gk\over n^2}\biggr[(\alpha_1-\alpha_2)+{k^2 T\over
  g(\rho_1+\rho_2)}\biggr]+1\biggr\}(\alpha_2 q_1+\alpha_1 q_2-k)\nonumber\\
&&-4k\alpha_1\alpha_2+ {4k^2\over n}(\alpha_1\nu_1-
  \alpha_2\nu_2)[\alpha_2q_1-\alpha_1q_2+k(\alpha_1-\nonumber\\
&&-\alpha_2)]+{4k^3\over n^2}(\alpha_1\nu_1-\alpha_2\nu_2)^2(q_1-k)(q_2-k)
= 0 .
\end{eqnarray}
The abbreviations $\alpha_{1,2}=\rho_{1,2}/(\rho_1+\rho_2)$ and 
$q_{1,2}^2=k^2+n/\nu_{1,2}$ were introduced where
$\nu_{1,2}=\mu_{1,2}/\rho_{1,2}$
is the kinematic viscosity for each region. For the rest
of the paper we will use a dimensionless surface tension
$S=T/\big[(\rho_1+\rho_2)$ $(g\nu_1^4)^{1/3}\big]$ if not stated otherwise.

We now briefly discuss the general results of the dispersion relation
(\ref{11}). The configuration where the lighter fluid is on top of the
heavier one, $\rho_2<\rho_1$, is always stable, i.e., Re$\{n(k)\}\leq 0$
for all $k$. This is independent of whether there is any surface tension
(Fig.\ \ref{fig71} (b)) or not (Fig.\ \ref{fig71} (a)). If the strata are in
the
opposite order, $\rho_2>\rho_1$, then the surface tension plays a
crucial role. In the case of no surface tension (Fig.\ \ref{fig72} (a))
the system is unstable against disturbances of any wave number, i.e.,
Re$\{n(k)\}\geq 0$ for all $k$. If there is a surface tension a critical
wave number exists
\begin{equation}
\label{12}
  k_c=\sqrt{{g\over T}(\rho_2-\rho_1)}\quad ,
\end{equation}
and the system is stable (unstable) against modes with wave numbers
which are larger (smaller) than $k_c$ (Fig.\ \ref{fig72} (b)).
A moderate variation in the relation between the two viscosities
$\nu_1$ and $\nu_2$ has no strong influence on the
general behavior of the growth rates (see Figs.\ \ref{fig71} and \ref{fig72}).

The experiments are carried out in a finite-size cell in contrast to our
simplification of infinite length in $z$ direction. We determine now the limits
to which this simplification is justified (see also \cite{mikaelian}).
If the walls of the cell are at $z=\pm L_z$ the appropriate 
ansatz for $w$ is
\begin{eqnarray}
\label{13}
w_1&=&a_1{\rm e}^{+kz}+a_2{\rm e}^{-kz}+b_1{\rm e}^{+q_1z}+b_2{\rm e}^{-q_1z}
  \quad{\rm ~for~}z\leq 0,\nonumber\\
&&\\
\label{14}
w_2&=&c_1{\rm e}^{+kz}+c_2{\rm e}^{-kz}+d_1{\rm e}^{+q_1z}+d_2{\rm e}^{-q_1z}
  \quad{\rm ~for~}z\geq 0.\nonumber\\
&&  
\end{eqnarray}
With the boundary condition $w=\partial_z w=0$ at $z=\pm L_z$ and the analysis
at the interface as above one ends up with a system of eight equations for the
constants in (\ref{13}-\ref{14}). The coefficients of the corresponding matrix
are given in the Appendix. The vanishing determinant leads to the dispersion
relation $n(k)$, its numerical solution is shown in Fig.\ \protect\ref{fig8} for
different lengths $L_z$. The comparison with the data for infinite $L_z$ shows
that there is no real difference as long as $|L_z|>3$ mm for the used material
parameters. Since the thickness of the sand layer is below this margin we
expect finite-size effects. In the case of asymmetrically arranged
walls at $z=-L_1$ and $z=+L_2$ the dispersion relation shows up to a ratio
of $|L_1|/L_2\simeq 17$ only a very weak deviation from the results for
symmetrical walls.

In the frame of our continuum approach we consider the water-sand mixture
as a suspension in accordance with the classification in \cite{rumpf}.
The dispersion medium is water and the dispersed material consists of
sand particles with a mean diameter and a density as stated in
Sect.~\ref{sec:2}.
The material density as well as the dynamical viscosity of the mixture depend
on the packing density $\phi$ of the granular material. The packing density
$\phi$ measures the volumetric concentration of the particles in the mixture.
The material density of the mixture is given by
\begin{equation}
\label{15}
   \rho_{mixture}=\phi\rho_{sand}+(1-\phi )\rho_{water}\quad .
\end{equation}
Since $\rho_{water}$ and $\rho_{sand}$ are constant, the considered small
disturbances in the mixture density $\delta\rho_{mixture}$ imply that $\phi$
varies according to Eq. (\ref{15}). In the following the index 1 refers to
water and 2 to the mixture, respectively. Two empirical formulae \cite{krieger,chong}
\begin{eqnarray}
\label{16}
{\mu_2\over \mu_1}=\mu_r&=&\left(1-{\phi\over \phi_{max}}\right)^{-2.5
\phi_{max}}\;\; \phi_{max}=0.63,\\
\label{17}
{\mu_2\over \mu_1}=\mu_r&=&\left(1+{0.75\over {\phi_{max}\over \phi}-1}
\right)^2\quad\;\;\;\; \phi_{max}=0.605,
\end{eqnarray}
were widely used for the dynamical viscosity of a hard sphere suspension.
The maximal packing densities in (\ref{16}, \ref{17}) result from the fit
of the proposed formulae to the experimental results.

Figure \ref{fig9} shows the behavior of the relative dynamical viscosity
$\mu_r$ for different packing densities $\phi$ according to Eqs.\ (\ref{16},
\ref{17}). A third relation, $\mu_r\simeq 1/\left[ 1-(\phi /\phi_{max})\right]
^{1/3}$ with $\phi_{max}=0.625$, was also plotted where
particle sizes beyond the colloidal range were incorporated into
the theoretical basis of this relation \cite{acrivos}. All three of them give
nearly the same value for $\mu_r (\phi )$ at moderately dense packings,
$0.4 \leq\phi\leq 0.48$. Above this region the relative viscosity diverges as
$\phi$ approaches its maximal value. But the strength of the divergent
behavior is remarkably different so that $\mu_r$ already varies about a factor
of 5 for $\phi =0.57$. The fact of wide spread measured values for the relative
viscosity of the same system is exemplarily shown in \cite{meeker}. The
measured values of $\mu_r$ vary between $21$ and $400$ for a rather colloid
system at
$\phi\approx 0.5$. A similar situation is present for suspensions \cite{ungarish}
which is why we cannot rely on a fixed value of $\mu_r$ within small error bars.
Therefore we consider $\mu_r$ a variable parameter within reasonable limits
rather than a fixed material parameter.

\section{Results and Discussion}
\label{sec:4}
In experiment I the height of the sedimented sand layer was measured by an
optical close-up with a resolution of $37$ $\mu$m per pixel. On the basis
of six independent samples the height was determined to $2.6\pm 0.2$ mm.
This results in a packing densitiy of $\phi=0.48\pm 0.04$.
Thus the mixture is sufficiently characterized by Eqs.\ (\ref{15}-\ref{17})
where $\rho_{water}=0.988\,{\rm g}/{\rm cm}^3$. By means of (\ref{17}) the
viscosity of the mixture is $\mu_2\simeq 15\mu_1$ for $\phi =0.48$. The
resulting growth rates show disagreement for infinite $L_z$ as well
as for finite $L_z$ (Fig.\ \ref{fig10}). Since the relative dynamical viscosity
is the most uncertain quantity in our calculations we vary $\mu_r$
to find the best fit with the experimental data. Under the assumption
of zero surface tension a least square fit results in $\mu_2\simeq 104\,\mu_1$
which gives a fairly good agreement over the whole $k$ range. In comparison to
$\mu_2\simeq 104\,\mu_1$, fits with $\mu_2\simeq 128\,\mu_1$ and
$\mu_2\simeq 85\,\mu_1$ show a better agreement for smaller and larger wave
numbers $k$, respectively.

For two sets of parameters, $\mu_2\simeq 15\,\mu_1$ and
$\mu_2\simeq 104\,\mu_1$, we calculate the dispersion relation for
infinite as well as for finite $L_z$. The finite-size effects appear
for small wave numbers and decrease with increasing wave numbers. The
differences in $n(k)$ for small wave numbers are more pronounced for
larger relative viscosities (see Fig.\ \ref{fig10}). The dispersion relations for
infinite and finite $L_z$ approach each other at wave numbers where the
initial amplitudes of the disturbances are in the order of $10^{-2}$ mm.
For these disturbances a boundary at $2.6$ mm distance appears
to be at infinity. Therefore it does not matter whether we choose
$L_z=\pm\infty$ or $L_z=\pm 2.6$ mm. This is not the case for small
wave numbers where the initial amplitudes are in the order of $10^{-1}$ mm
(Fig.\ \ref{wavenumber} (b)).

There is one viscosity measurement \cite{arnaud} which comes near to the value
of the relative viscosity suggested by our fit. The measurement was
carried out for crushed sand with particle diameters from $20$ to $80$ $\mu$m
and gives a value of $\mu_2\sim 110\,\mu_1$ for $\phi =0.48$ (see Fig. 3 in
\cite{arnaud}). This value for the relative viscosity is notably close to
our fit value. The values for the viscosity of the suspension estimated in
\cite{didwania}, $\mu_2\sim (130-190)\,\mu_1$, are in a similar range as
ours. These values are determined by the help of the
{\it singular} wave number with the
largest growth rate. A real comparison with our fit value is not possible
because the necessary packing density is not given for the type of
experiments from which the viscosity values were estimated. 

Nevertheless, the large difference in $\mu_r$ between
\cite{krieger,chong,acrivos} and \cite{arnaud} over a wide range of the
packing density $\phi$ represents an unsatisfying situation. It highlights
the need for comprehensive and unambiguous viscosity measurements in highly
concentrated hard sphere suspensions. It further shows how
sensitive $\mu_r$ is on experimental methods \cite{acrivos}, the accurate
determination of the density packing \cite{meeker}, and the type of flow
involved in the measurements \cite{paetzold}.

The aspect ratio of the cell suggests that a description refering to the
Hele-Shaw type of the cell might be closer to the experimental
configuration. Adapting the dispersion relation of the Saffman-Taylor
instability to the case of zero throughflow velocity leads to
\cite{saffman}
\begin{equation}
\label{18}
n(k)={b^2\over 12(\mu_1 +\mu_2)}\left[k g(\rho_2-\rho_1)- T k^3\right]
\end{equation}
where $b=4$ mm denotes the width of the cell. In contrast to the Rayleigh-Taylor
approach Eq.\ (\ref{18}) contains two parameters, the dynamical viscosity
$\mu_2$ of the mixture and the surface tension $T$, which have to be determined
by a least square fit. The validity of the Hele-Shaw approach is limited by a
cut-off condition at which the wave number exceeds $2\pi/b$.

The fit values $\mu_2\simeq 404\,\mu_1$ and $T\simeq 1.06\cdot 10^{-3}$
Nm$^{-1}$ result in a fit curve which is inferior to the Rayleigh-Taylor fit
(Fig.\ \ref{fig11}). Additionally, both fit parameters are questionable. The value
$\mu_2\simeq 404\,\mu_1$ is beyond any realistic one for the dynamical viscosity
of the mixture at a packing density of $\phi=0.48$. A nonzero surface
tension between the water-sand mixture and water is also arguable.

The better agreement of the Rayleigh-Taylor approach with the experimental data
is backed up by calculations focused on the influence of finite cell boundaries
for Rayleigh-B\'enard convection \cite{frick}. As long as the ratio between the
relevant height of the cell and its width is smaller than 0.8, the assumption
of infinite boundaries in $y$ direction is a good approximation. If this ratio
is larger than 5 the Hele-Shaw approach is well-founded. The height of the sand
layer and the width of the cell give a ratio of 0.65 which supports the
Rayleigh-Taylor approach. As a consequence the cell width has no influence
on the instability for the used material sets.

It has to be stressed that both the Hele-Shaw and the Rayleigh-Taylor
approach assume a trivial $y$ dependence of the flow. In the experiment,
that assumption is not totally fulfiled: The rotation of the apparatus leads to a
sand layer which is not perfectly flat even in the beginning of the flow
process. This will lead to a three-dimensional flow which effects the
appearing patterns and their wavelengths. The strength of this effect is,
however, presently hard to determine.

\section {Concluding Remarks}
\label{sec:5}
In a closed Hele-Shaw-like cell the temporal evolution of
a water-sand interface was investigated. For the unstable
stratification, sand above water, the instability is driven by gravity.
The images of the temporal evolution were analyzed by DFT. The Fourier
spectra show that the initial disturbances of the interface grow
exponentially at the beginning of the pattern forming process.
This enables us to determine the growth rates by an exponential fit for
every wave number in our spectra. The data show that the growth rate increases
with increasing wave number until it saturates at larger values of
$k$. This general behavior is not influenced by the mean particle diameter
which was tested with three different size distributions. Experiments
with different amounts of sand reveal a shift of the dispersion relation
$n(k)$ as a whole. By increasing the mass of sand, $n(k)$ is shifted towards
larger values of $n$.

To describe the general behavior we choose a two-fluid
system as a model. Carrying out a linear stability analysis for the
interface between the two fluids we calculate the growth rates from the
dispersion relation for a finite-size cell. The theoretical results agree
with the essence in the experimental findings when assuming a
relative viscosity of the water-sand mixture which is close to one measured
with crushed sand and water \cite{arnaud}.

Considering our simplifications and the uncertainty in one relevant material
parameter, the continuum approach gives a reasonable agreement with the
experimental results. An interpretation for such an agreement was proposed
in \cite{didwania}. Above a critical value of the packing density the mixture
exhibits a non-zero yield stress well known for dry granular material. Below
this critical value the yield stress vanishes and the mixture behaves more
like a fluid with effective properties. Applying this interpretation, the
measured packing density of $\phi=0.48$ is below this critical value. Thus,
our results confirm a certain analogy between concentrated
suspensions and fluids which was found also in numerical simulations
\cite{druzhinin}. However, further investigations need to be done in order
to clarify open questions.

The available experimental data give no hint on whether the dispersion relation
$n(k)$ will have a second zero at $k_c$ after the observed plateau.
A critical wave number $k_c$ means a non-zero interfacial tension according to
(\ref{12}).
A surface tension acts in a way to minimize the surface of the fluid.
Therefore it suppresses the formation of waves with large wave numbers
because their creation entail additional surface. The suppression leads to
a saturation (reduction) in the growth rates of disturbances with large
(very large) wave numbers. In the context of miscible fluids with slow diffusion
the concept of an effective dynamical surface tension was recently successfully
applied \cite{lopez}. The reason for such a surface tension lies in the attraction
between moving particles in a fluid for non-zero Reynolds numbers \cite{jayaweera}.
The attraction originates from the dynamics in the viscous fluid and results
in a favoured distance between the moving particles. Consequently, is costs
energy to separate the particles beyond this favoured distance, i.e., to dilute the
suspension. The necessary energy corresponds to a surface tension which
is called effective {\it dynamical} surface tension to emphasize its dynamical
origin. The effect of such an effective dynamical surface tension is seen in
experiments by the presence of sharp interfaces in rising bubbles, falling
drops \cite{joseph} and of growing deposits \cite{lopez}.
A concentration gradient concerning sand is present in our system between
the water and the water-sand layer. By reducing the density difference
between the layers, measurements for $k$
values beyond the plateau should be possible. This may lead towards the
determination of $k_c$ and the interfacial tension, respectively. 

Since a linear stability analysis is restricted to terms linear in
the disturbances we cannot take into account fluctuations in the relative
viscosity of the mixture. These would lead to terms of higher
order because all terms which contain the viscosity are already linear
in the velocity disturbances (see {\ref{1}-\ref{3}). On the other hand,
considering fluctuations of $\rho_{mixture}(\phi)$, i.e., fluctuations of the
packing density $\phi$, implies variations of $\mu_r (\phi )$.
This inconsistency owing to the restrictions of our linear theory
can only be resolved in a nonlinear analysis.

\begin{acknowledgement}
\label{sec:6}
We are grateful to Lluis Carrillo, Stefan Neser, and Stefan Schwarzer for
inspiring discussions. The experiments were supported by DFG through
Re 588/11-1.
\end{acknowledgement}

\section{Appendix}
\label{sec:7}
The matrix elements for finite $L_z$ are
  $A_{11}={\rm e}^{-kL_z}$, $A_{12}={\rm e}^{kL_z}$, $A_{13}={\rm e}^{-q_1L_z}$,
$A_{14}={\rm e}^{q_1L_z}$, $A_{15}=A_{16}=A_{17}=A_{18}=0$,
  $A_{21}=k{\rm e}^{-kL_z}$, $A_{22}=-k{\rm e}^{kL_z}$,
$A_{23}=q_1{\rm e}^{-q_1L_z}$, $A_{24}=-q_1{\rm e}^{q_1L_z}$,
$A_{25}=A_{26}=A_{27}=A_{28}=
  A_{31}=A_{32}=A_{33}=A_{34}=0$, $A_{35}={\rm e}^{kL_z}$,
$A_{36}={\rm e}^{-kL_z}$, $A_{37}={\rm e}^{q_2L_z}$, $A_{38}={\rm e}^{-q_2L_z}$,
  $A_{41}=A_{42}=A_{43}=A_{44}=0$, $A_{45}=k{\rm e}^{kL_z}$,
$A_{46}=-k{\rm e}^{-kL_z}$, $A_{47}=q_2{\rm e}^{q_2L_z}$,
$A_{48}=-q_2{\rm e}^{-q_2L_z}$,
  $A_{51}=A_{52}=A_{53}=A_{54}=1$, $A_{55}=A_{56}=A_{57}=A_{58}=-1$,
  $A_{61}=k$, $A_{62}=-k$, $A_{63}=q_1$, $A_{64}=-q_1$, $A_{65}=-k$, $A_{66}=k$,
$A_{67}=-q_2$, $A_{68}=q_2$,
  $A_{71}=A_{72}=2\mu_1k^2$, $A_{73}=A_{74}=\mu_1(q_1^2+k^2)$,
$A_{75}=A_{76}=-2\mu_2k^2$, $A_{77}=A_{78}=-\mu_2(q_1^2+k^2)$,
  $A_{81}=R/2-C-\alpha_1$, $A_{82}=R/2+C+\alpha_1$, $A_{83}=R/2-q_1C/k$,
$A_{84}=R/2+q_1C/k$, $A_{85}=R/2-C+\alpha_2$, $A_{86}=R/2+C-\alpha_2$,
$A_{87}=R/2-q_2C/k$, and $A_{88}=R/2+q_2C/k$. Furthermore, the abbreviations
\begin{eqnarray}
\label{A1}
  R&=&{k\over n^2}\left[g(\alpha_2-\alpha_1)+{k^2T\over \rho_1+\rho_2}
  \right]\quad{\rm ~and}\\
\label{A2}
  C&=&{k^2\over n}(\alpha_1\nu_1-\alpha_2\nu_2)
\end{eqnarray}
were used.

\begin{figure}[htp]
   \includegraphics[width=8.4 cm]
   {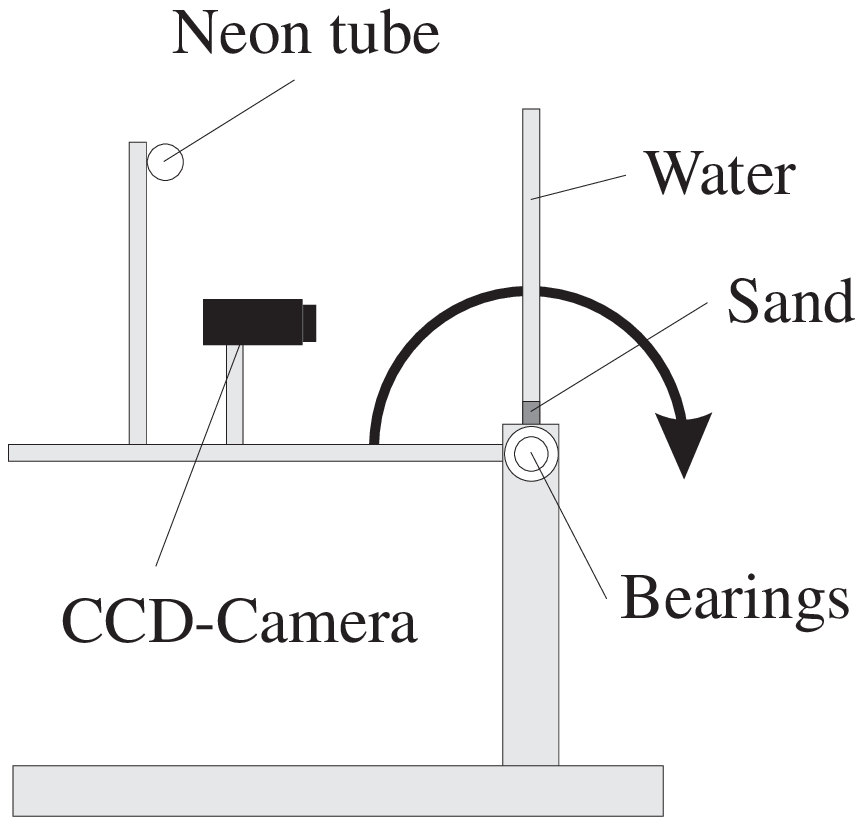}
   \caption{Experimental setup.}
   \label{setup}
\end{figure}

\begin{figure}[htp]
  \includegraphics[width=8.6 cm]
  {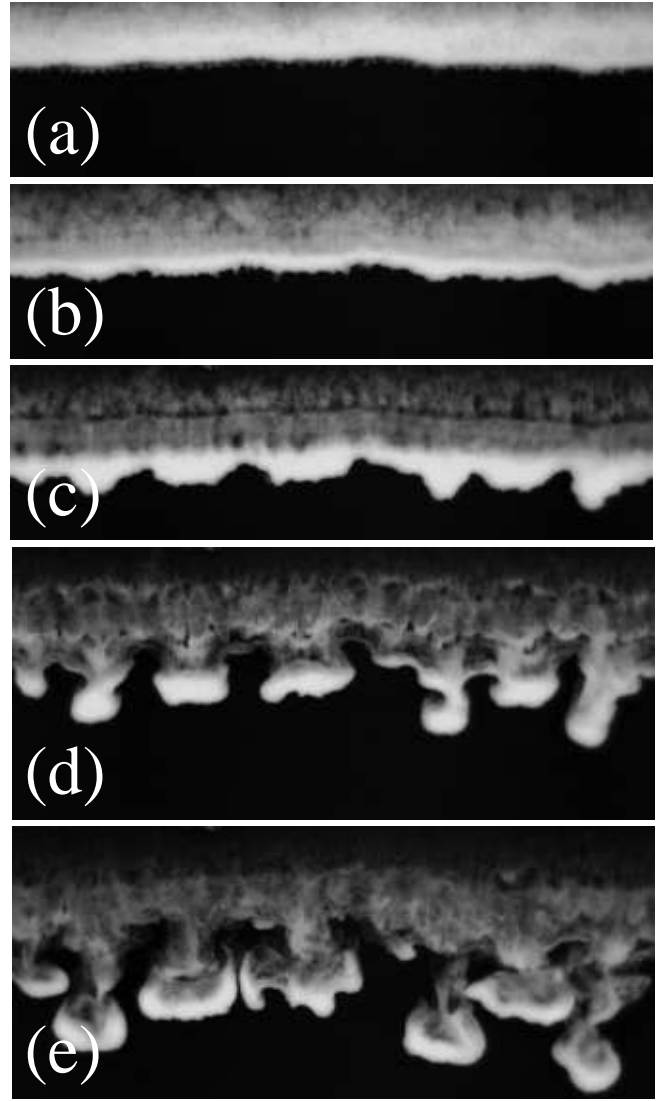}
  \caption{Sand-water interfaces at certain time steps: (a) 20 ms, (b) 80 ms,
  (c) 140 ms, (d) 200 ms, and (e) 260 ms. The size distribution of the sand
  particles is given by 71-80 $\mu$m and the sand mass in the cell by 2 g,
  respectively. The presented frames show the middle part of the cell and have
  a horizontal length of 68 mm.}
  \label{pattern}
\end{figure}

\begin{figure}[htp]
  \includegraphics[width=8.6 cm]
  {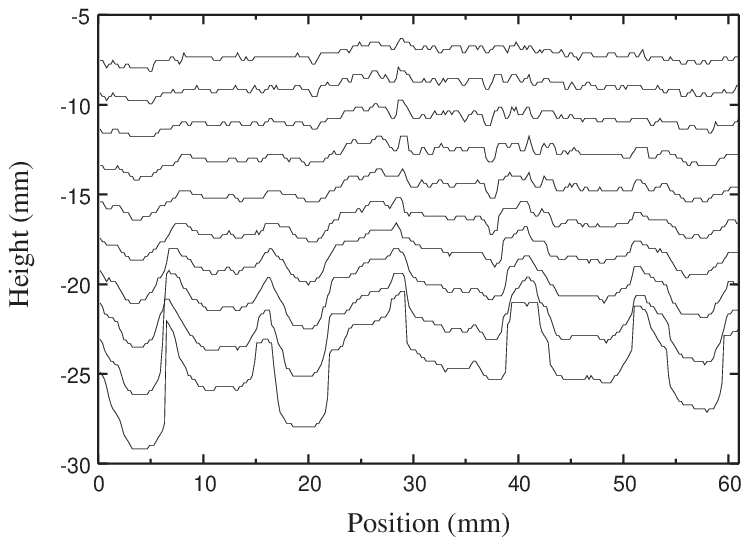}
  \caption{Temporal evolution of the water-sand interfaces. The patterns are
  detected every 20 ms and shown with a constant vertical offset of 1 mm. The
  experimental conditions are the same as in Fig.\ \protect\ref{pattern}.}
  \label{temporal}
\end{figure}

\begin{figure}[htp]
  \includegraphics[width=8.6 cm]
  {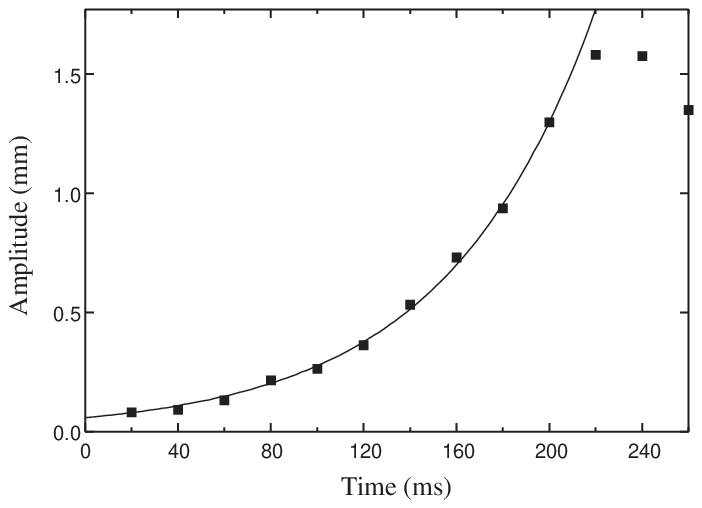}
  \caption{Amplitude $A$ of a DFT-analysis for a typical wave number (here
  $k$ = 7 cm$^{-1}$) in dependence on time $t$.  An exponential fit is obtained
  by Eq.\ (\protect\ref{efit}). The values belong to the interfaces presented in
  Fig.\ \protect\ref{temporal}.}
  \label{fit}
\end{figure}

\begin{figure}[htp]
  \includegraphics[width=8.0 cm]
  {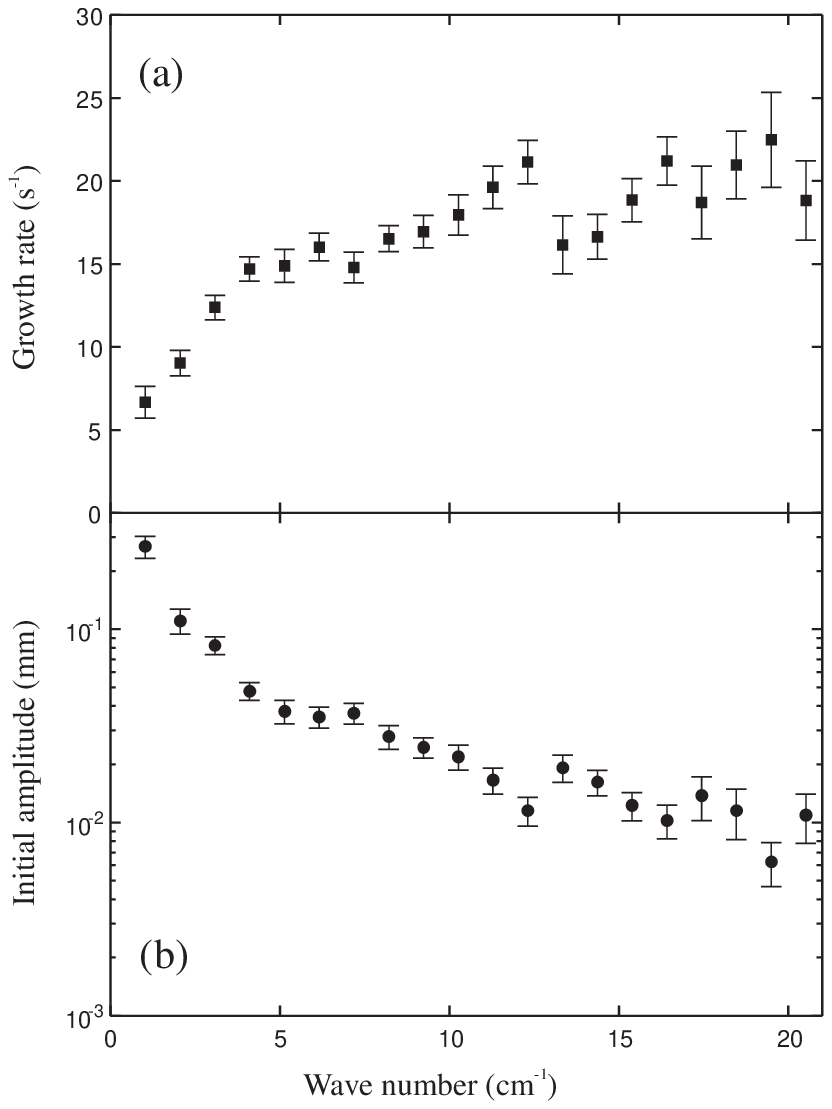}
  \caption{Growth rate (a) and initial amplitude (b) versus the
  wave number $k$ for experiment I (see Table \protect\ref{table1}).}
  \label{wavenumber}
\end{figure}

\begin{figure}[htp]
  \includegraphics[width=8.4 cm]
  {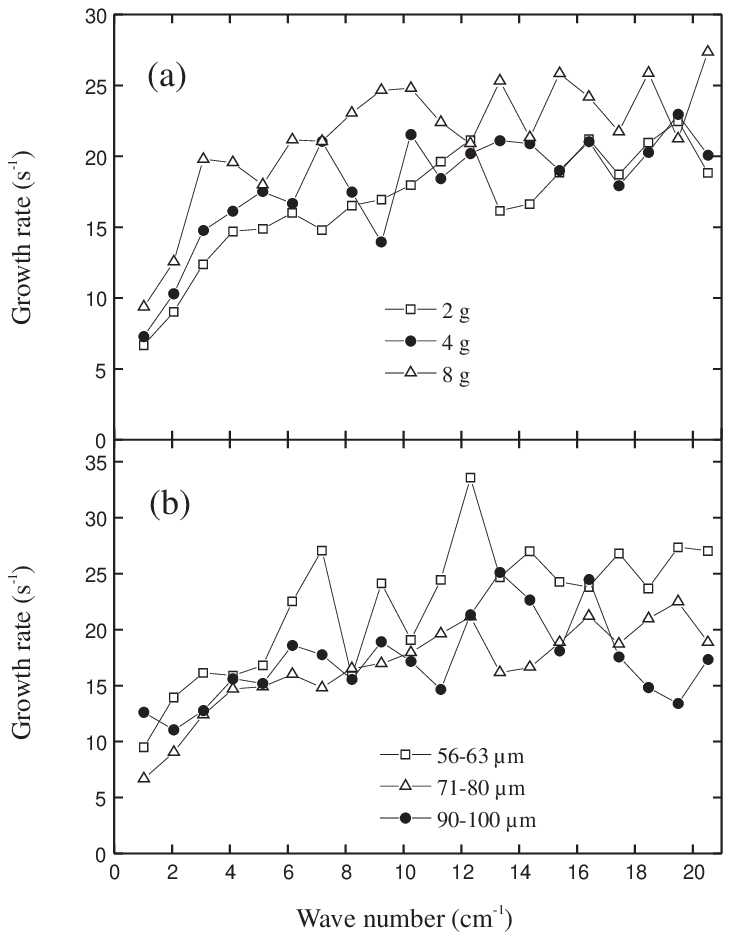}
  \caption{Growth rate versus the wave number for different material parameters.
  In (a) 2 g ($\Box$), 4 g ($\bullet$), and 8 g ($\scriptstyle{\triangle}$) of
  sand with the same size distribution were used (Experiment I, II, III). In (b)
  2 g of sand with size distributions of $56 - 63$ $\mu$m ($\Box$),
  $71 - 80$ $\mu$m ($\scriptstyle{\triangle}$), and $90 - 100$ $\mu$m
  ($\bullet$) were used (Experiment IV, I, V).}
  \label{material}
\end{figure}

\begin{figure}[htp]
  \includegraphics[width=8.3 cm]
  {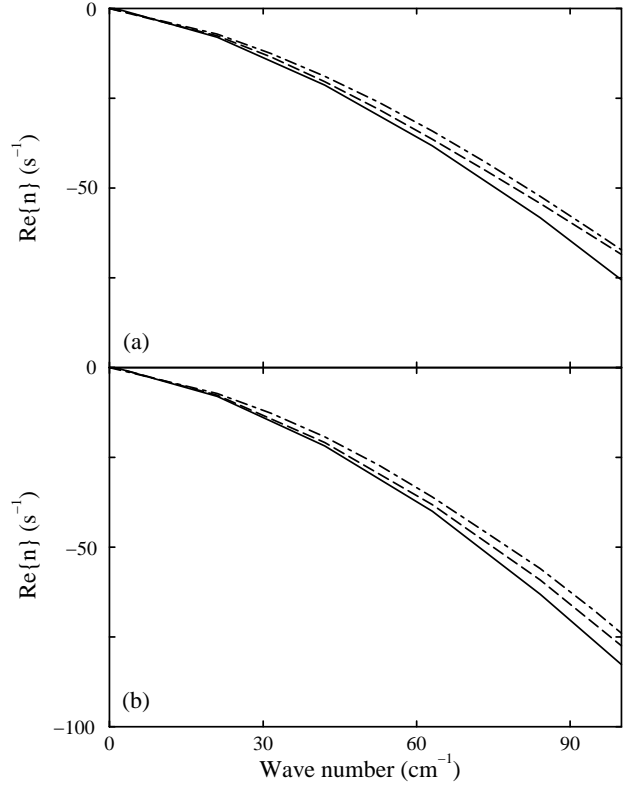}
  \caption{For zero surface tension (a) and surface tension $S=1$ (b)
  the growth rate Re$\{n\}$ of the modes is plotted versus their wave
  number $k$ for the case that the top fluid is lighter than the bottom one,
  $\rho_2=0.5\rho_1$. The surface tension as well as the relation of the
  viscosities does not change the {\it overall} behavior too much as the three
  examples show:
  $\nu_1=\nu_2=10^{-6}\,{\rm m}^2\,{\rm s}^{-1}$ (solid line),
  $\nu_1=10^{-6}\,{\rm m}^2\,{\rm s}^{-1}$ and $\nu_2=0.75\nu_1$ (long-dashed
  line), and $\nu_2=10^{-6}\,{\rm m}^2\,{\rm s}^{-1}$ and $\nu_1=0.75\nu_2$
  (dot-dashed line).}
  \label{fig71}
\end{figure}

\begin{figure}[htp]
  \includegraphics[width=8.3 cm]
  {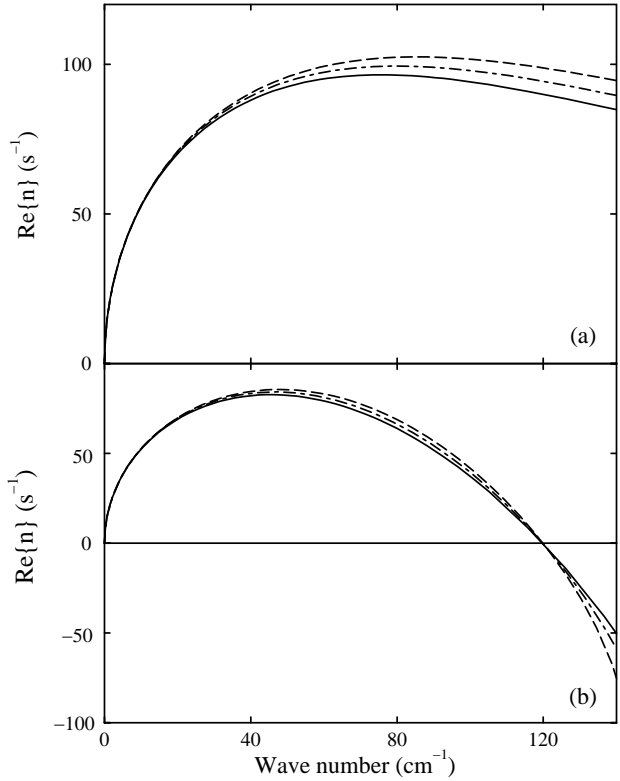}
  \caption{The growth rate Re$\{n\}$ of the modes is plotted versus their wave
  number $k$ for the case that the top fluid is heavier than the bottom
  one, $\rho_2=2\rho_1$. The surface tension $S$ is zero in (a) and $1$ in (b).
  The surface tension causes a drastic change in the
  behavior whereas the ratio of the viscosities does not change the
  {\it overall} behavior too much as the three examples show:
  $\nu_1=\nu_2=10^{-6}\,{\rm m}^2\,{\rm s}^{-1}$ (solid line),
  $\nu_1=10^{-6}\,{\rm m}^2\,{\rm s}^{-1}$ and $\nu_2=0.75\nu_1$ (long-dashed
  line), and $\nu_2=10^{-6}\,{\rm m}^2\,{\rm s}^{-1}$ and $\nu_1=0.75\nu_2$
  (dot-dashed line).}
  \label{fig72}
\end{figure}

\begin{figure}[htp]
  \includegraphics[width=8.3cm]
  {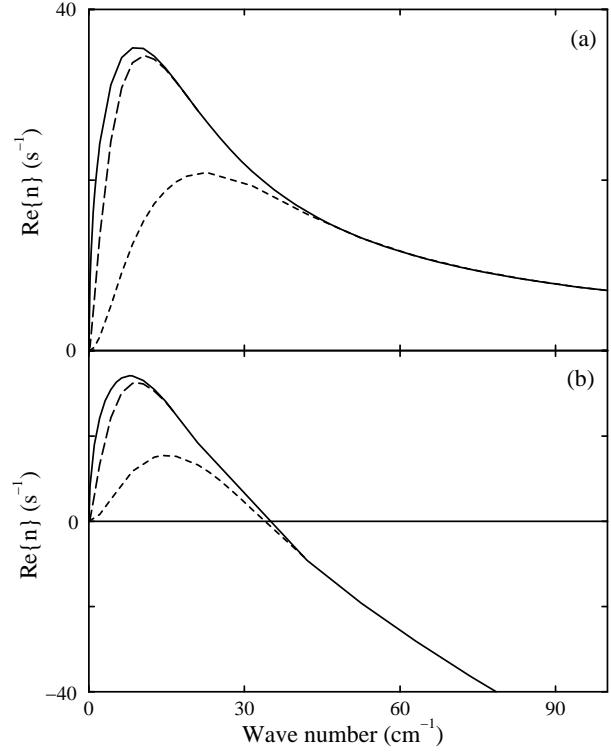}
  \caption{For zero surface tension (a) and surface tension $S=12.8$ (b) the
  growth rate Re$\{n\}$ is plotted versus the wave number $k$ for
  $L_z=\pm3\,$mm (long-dashed line) and $L_z= \pm 1\,$mm (dashed line). For
  comparison the solid line shows the graph for infinite $L_z$. It reveals that
  finite-size effects play a significant role if $k< 2\pi/|L_z|$. The material
  parameters are $\rho_2 =2\rho_1$ and $\nu_2=33\nu_1$ with
  $\rho_1=1\,{\rm g}\,{\rm cm}^{-3}$ and
  $\nu_1=10^{-6}\,{\rm m}^2\,{\rm s}^{-1}$.}
  \label{fig8}
\end{figure}

\begin{figure}[htp]
  \includegraphics[width=8.3 cm, height=7.0 cm]
  {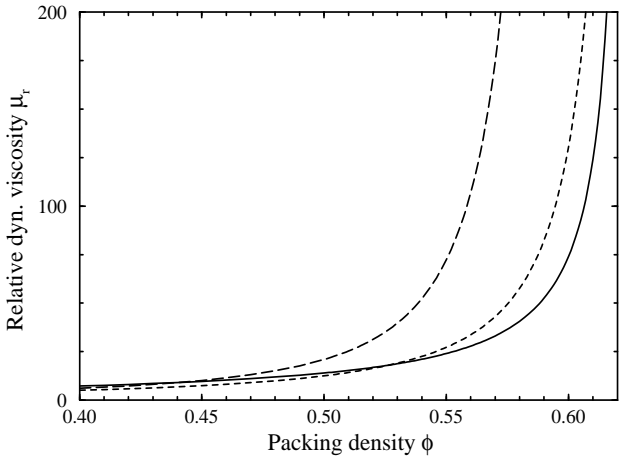}
  \caption{Relative dynamical viscosity $\mu_r$ versus the packing density
  $\phi$. The results of the two empirical formulae (\ref{16}, \ref{17}) are
  plotted as dashed and long-dashed lines, respectively. The solid line indicates
  $\mu_r\simeq 1/\left[ 1-(\phi /\phi_{max})\right]^{1/3}$ with $\phi_{max}=0.625$
  \protect\cite{acrivos}. $\mu_r$ is nearly the same for all three approaches
  provided the packing is not too dense, $\phi\leq 0.48$. Above this range
  $\mu_r$ starts to diverge as $\phi$ reaches $\phi_{max}$ where the
  divergent behavior differs significantly between the various approximations.
  }
  \label{fig9}
\end{figure}

\begin{figure}[htp]
  \includegraphics[width=8.3 cm, height=7.0 cm]
  {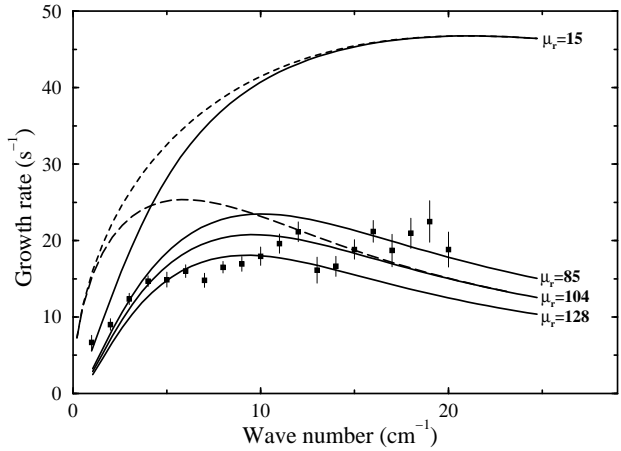}
  \caption{Growth rates $n$ of the disturbances against their wave numbers $k$
  for a packing density $\phi =0.48$. The results $n(k)$ for infinite $L_z$
  according to (\ref{11}) are shown by the dashed line ($\mu_r =15$) and the
  long-dashed line ($\mu_r =104$), respectively. The solid lines are calculated
  with $|L_z|=\pm 2.6\,{\rm mm}$, the measured height of the sand layer. The
  relative viscosity for each drawn curve is stated at the right end of it.
  The finite-size effects appear for small wave numbers and increase with
  increasing relative viscosity. A relative viscosity of
  $\mu_r =15$ based on (\protect\ref{17}) gives growth rates which are far
  away from the experimental results (\protect\rule{1.0mm}{1.0mm}).
  The best fit over the whole $k$ range gives $\mu_r=104$.
  Fits with slightly smaller or larger $\mu_r$ values deliver better agreements
  either with larger or smaller wave numbers.}
  \label{fig10}
\end{figure}

\begin{figure}[htp]
  \includegraphics[width=8.3 cm, height=7.0cm]
  {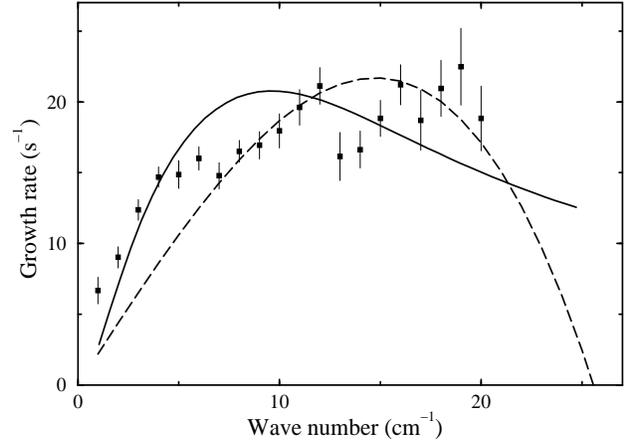}
  \caption{Growth rates $n$ of the disturbances against their wave numbers $k$
  for a packing density $\phi =0.48$. Two different fits are compared with
  the experimental results (\protect\rule{1.0mm}{1.0mm}). The solid line
  shows the graph based on the Rayleigh-Taylor approach with finite $z$
  boundaries at $|L_z|=\pm 2.6\,{\rm mm}$ and $\mu_r =104$. The long-dashed
  line presents the results of the Hele-Shaw approach with
  $\mu_r =404$ and $T\simeq 1.06\cdot 10^{-3}$ Nm$^{-1}$.
  }
  \label{fig11}
\end{figure}

\begin{table}
\caption{Details of sand used in certain experimental configurations.}
\label{table1}
\begin{tabular}{cccc}
\hline\noalign{\smallskip}
Experiment&Size distribution [$\mu m$]&Mass [g]&Runs  \\
\noalign{\smallskip}\hline\noalign{\smallskip}
I&71-80&2&100\\
II&71-80&4&12\\
III&71-80&8&12\\
IV&56-63&2&12\\
V&90-100&2&12\\
\noalign{\smallskip}\hline
\end{tabular}
\end{table}

\end{document}